\documentstyle[]{article}
\begin{document}
\title{Self-Diffusion by Multivariate-Normal Turbulent Velocity Field}
\author{R.~V.~R.~Pandya\thanks{Email address: {\sl
rvrpturb@uprm.edu}}, \\Department of Mechanical Engineering,\\
University of Puerto Rico at Mayaguez, Mayaguez,
PR 00680, USA }

\date{\today}
\maketitle

\begin{abstract}

A closed set of \textit{exact} equations describing statistical
theory of turbulent self-diffusion by multivariate-normal turbulent
velocity field is derived. In doing so, we first suggest exact
formulas for correlations $\langle f_i(p)f_j(p')R[{\bf f}] \rangle$,
$\langle g(p)R[{\bf f}] \rangle$ and $\langle g(p)f_j(p')R[{\bf f}]
\rangle$ when the functional $R[{\bf f}]$ is functional of functions
${f_i}$'s having multivariate-normal distribution, $g$ and $f_i$'s
have joint normal distribution and zero mean values.

\end{abstract}

Originative works of Taylor \cite{Taylor21} and Batchelor
\cite{Batchelor49} on turbulent self-diffusion of fluid particles
in Lagrangian and Eulerian frameworks, respectively, have placed
statistics of Lagrangian velocity field, first put forward by
Taylor \cite{Taylor21}, as fundamental properties in the field of
statistical theory of diffusion. Despite many persistent efforts
(for review, see e.g., Refs.
\cite{MY75,McComb90}), the theory of diffusion remains incomplete
mainly due to involved closure problems
\cite{McComb90,Roberts61,Kraichnan66} and lack of accurate
prediction of Lagrangian statistics
\cite{Corrsin59,Kraichnan70,Saffman62} in general turbulent flows.
In this letter, we base our Eulerian analysis on the equation for
a passive scalar field $\psi({\bf x},t)$ whose ensemble average
(denoted by $\langle \,\rangle$) denotes Green's function $G$ for
the evolution in space-time (${\bf x}-t$) of an arbitrary initial
probability distribution for the position of marked fluid
particles in space \cite{Batchelor49,Roberts61,Kraichnan70}. We
solve the closure problem arising in the equation for $\langle
\psi({\bf x},t)\rangle$ resulting in an appearance of Lagrangian
velocity correlations. We then obtain equations relating
Lagrangian and Eulerian correlations for the velocity field. These
equations are exact when the fluctuations $u_i'$ in turbulent
velocity field $u_i({\bf x},t)=U_i({\bf x},t)+u_i'({\bf x},t)$
over the mean velocity $U_i({\bf x},t)$ obey multivariate-normal
distribution. In doing so, we first suggest three new functional
formulas which along with the Furutsu-Donsker-Novikov functional
formula
\cite{Furutsu63} are used in
our Eulerian analysis. These four functional formulas are now
discussed briefly before presenting the Eulerian analysis.

\textit{Functional Formulas:} Furutsu-Donsker-Novikov formula
\cite{Furutsu63} suggests that for any random function $f_i(p)$
with normal or Gaussian distribution and $\langle
f_i(p)\rangle=0$, its correlation with functional $R[{\bf f}]$ of
function $f_i$'s can be exactly given by
\begin{equation}
\langle f_i(p)R[{\bf f}] \rangle =\int \langle f_i(p)f_k(p')
\rangle \left \langle\frac{\delta R[{\bf f}]}{\delta f_k(p')dp'}
\right \rangle dp', \label{N1}
\end{equation}
%
where the integral extends over the region of arguments $p$ in
which the function $f_i$ is defined and $  \left
\langle\frac{\delta R[{\bf f}]}{\delta f_k(p')dp'} \right \rangle$
represents the functional derivative of $R$ with respect to $f_k$.
Here, we suggest another exact functional formula for higher order
correlations $\langle f_i(p)f_j(p')R[{\bf f}]\rangle$, written as
\begin{eqnarray}
\langle f_i(p)f_j(p')R[{\bf f}]\rangle =&& \langle
f_i(p)f_j(p')\rangle \langle R[{\bf f}]\rangle+\frac{1}{2}\int
\langle f_i(p)f_k(s) \rangle \left \langle f_j(p')\frac{\delta
R[{\bf f}]}{\delta f_k(s)ds} \right \rangle ds \nonumber \\
&&+\frac{1}{2}\int \langle f_j(p')f_k(s) \rangle \left \langle
f_i(p)\frac{\delta R[{\bf f}]}{\delta f_k(s)ds} \right \rangle ds.
\label{formula2}
\end{eqnarray}
Yet another formulas which can be obtained from Eqs. (\ref{N1})
and (\ref{formula2}), respectively, are
\begin{equation}
\langle g(p)R[{\bf f}] \rangle =\int \langle g(p)f_k(p') \rangle
\left \langle\frac{\delta R[{\bf f}]}{\delta f_k(p')dp'} \right
\rangle dp', \label{N3}
\end{equation}
\begin{eqnarray}
\langle g(p)f_j(p')R[{\bf f}]\rangle =&& \langle
g(p)f_j(p')\rangle \langle R[{\bf f}]\rangle+\frac{1}{2}\int
\langle g(p)f_k(s) \rangle \left \langle f_j(p')\frac{\delta
R[{\bf f}]}{\delta f_k(s)ds} \right \rangle ds \nonumber \\
&&+\frac{1}{2}\int \langle f_j(p')f_k(s) \rangle \left \langle
g(p)\frac{\delta R[{\bf f}]}{\delta f_k(s)ds} \right \rangle ds
\label{N4}
\end{eqnarray}
where $g$ and $f_i$'s have joint normal distribution and $\langle
g\rangle =0$. It should be noted that the terms with functional
derivatives in Eqs. (\ref{formula2}) and (\ref{N4}) can be further
simplified by using Eqs. (\ref{N1}) and (\ref{N3}). From now
onward, these functional formulas given by Eqs. (\ref{N1}),
(\ref{formula2}), (\ref{N3}) and (\ref{N4}) are referred to as
$F1$, $F2$, $F3$ and $F4$, respectively.

Now, we briefly describe equations which can be used for
verifications of the functional formulas. Using Novikov's notation,
expansion of $R[{\bf f}]$ around ${\bf f}={\bf 0}$ can be written as
\begin{equation}
R[{\bf f}]=R[{\bf 0}]+\sum_{n=1}^\infty \frac{1}{n!}\int \cdots
\int R^{(n)}_{i_1\cdots i_n}(s_1,\cdot,s_n)f_{i_1}(s_1)\cdots
f_{i_n}(s_n)ds_1\cdots ds_n \label{func1}
\end{equation}
where
\begin{equation}
R^{(n)}_{i_1\cdots i_n}(s_1,\cdot,s_n)=\frac{\delta^nR[{\bf
f}]}{\delta f_{i_1}(s_1)ds_1\cdots\delta f_{i_n}(s_n)ds_n}
\,\Big|_{{\bf f}={\bf 0}} \label{func2}
\end{equation}
and summation over the repeated indices $i_1,i_2 \cdots i_n$ is
taken. The functional derivative of Eq. (\ref{func1}) yields
\begin{equation}
\frac{\delta R[{\bf f}]}{\delta f_k(s)ds}=\frac{\delta R[{\bf f}]
}{\delta f_k(s)ds}\Big|_{{\bf f}={\bf
0}}+\sum_{n=2}^{\infty}\frac{1}{(n-1)!}\int\cdots \int
R^{(n)}_{ki_{2}\cdots i_n}(s,s_2,\cdots,s_n)f_{i_2}(s_2)\cdots
f_{i_n}(s_n)ds_2\cdots ds_n. \label{func3}
\end{equation}
For multivariate-normal distribution for $f_i$'s and joint normal
distribution for $g$ and $f_i$'s
\begin{equation}
\langle A(s)f_{i_1}(s_1)\cdots f_{i_m}(s_m)\rangle= 0,
\end{equation}
\begin{equation}
\langle A(s)f_{i_1}(s_1)\cdots
f_{i_n}(s_n)\rangle=\sum_{\alpha=1}^n\langle
A(s)f_{i_\alpha}(s_\alpha)\rangle \langle f_{i_1}(s_1)\cdots
f_{i_{\alpha-1}}(s_{\alpha-1})f_{i_{\alpha+1}}(s_{\alpha+1})\cdots
f_{i_n}(s_n) \rangle \label{func4}
\end{equation}
where $A(s)$ is equal to any $f_i(s)$ or $g(s)$ and $m$ is an even
integer. Using Eqs. (\ref{func1})-(\ref{func4}), formulas $F2$,
$F3$ and $F4$ can be verified.

\textit{Eulerian Analysis}: We consider scalar field
\begin{equation}
\psi ({\bf x},t)=\delta({\bf x}-{\bf X(t)})
\end{equation}
for the Eulerian description of the self-diffusion of fluid
particle whose position ${\bf X}(t)$ (also denoted by components
$X_i(t)$) at any time $t$ is governed by the Lagrangian equation
\begin{equation}
\frac{dX_i(t)}{dt}=u_i({\bf x}={\bf X}(t),t)
\end{equation}
with initial condition ${\bf X}(t_0)={\bf x}_0$ at time $t_0$. The
governing equation for $\psi$ can be written as (see e.g.
\cite{Roberts61})
\begin{equation}
\frac{\partial }{\partial t}\psi({\bf
x},t)+\frac{\partial}{\partial x_i}[\psi({\bf x},t)u_i({\bf
x},t)]=0, \,\,\forall \,t > t_0 \label{psi1}
\end{equation}
with initial condition $\psi({\bf x},t_0)=\delta ({\bf x}-{\bf
x_0})$. The ensemble average of $\psi$ multiplied by $d{\bf x}$,
i.e. $\langle \psi\rangle d{\bf x}\equiv G({\bf x,t}|{\bf
x}_0,t_0)d{\bf x}$, with the Green's function $G$, denotes the
probability that a fluid particle at $({\bf x}_0, t_0)$ should lie
within a volume $d{\bf x}$ at later time $t > t_0$
\cite{Roberts61}. The ensemble average of Eq. (\ref{psi1})
\begin{equation}
\frac{\partial }{\partial t}\langle \psi({\bf
x},t)\rangle+\frac{\partial}{\partial x_i}[\langle \psi\rangle
U_i({\bf x},t)+\langle \psi u_i'({\bf x},t) \rangle]=0
\label{psi2}
\end{equation}
with initial condition $\langle \psi({\bf x},t_0)\rangle=\delta
({\bf x}-{\bf x_0})$ poses closure problem due to unknown
correlation $\langle \psi({\bf x},t) u_i'({\bf x},t) \rangle$. In
our analysis, we would need yet another scalar function
$\psi_b({\bf y},t-s)=\delta({\bf y}-{\bf X}(t-s))$ with initial
condition given at time $t$ i.e. $\psi_b({\bf
y},t)=\langle\psi_b({\bf y},t)\rangle=\delta({\bf y}-{\bf
X}(t))=\delta({\bf y}-{\bf x})$. In fact, $\langle\psi_b({\bf
y},t-s)\rangle d{\bf y}$ with $s > 0$ represents the probability
that a fluid particle lies within a volume $d{\bf y}$ at $({\bf
y}, t-s)$ if the particle reaches $({\bf x},t)$. The governing
equation for $\psi_b({\bf y}, t-s)$, after introducing notation
$s^-\equiv t-s$, is
\begin{equation}
\frac{\partial }{\partial s}\psi_b({\bf
y},s^-)-\frac{\partial}{\partial y_i}[\psi_b({\bf y},s^-)u_i({\bf
y},s^-)]=0,\label{psib}
\end{equation}
which is obtained by using the Lagrangian equation
\begin{equation}
-\frac{d}{ds}X_i(s^-)=u_i({\bf y}={\bf X}(s^-),s^-), \,\forall\,
0\le s \le t-t_0
\end{equation}
describing the backward evolution of trajectory  of fluid particle
with known initial position ${\bf X}(t)$ at initial time $s^-=t$
or $s=0$. The ensemble average of Eq. (\ref{psib}) poses closure
problem due to the unknown correlation $\langle\psi_b({\bf
y},s^-)u_i'({\bf y},s^-)\rangle$.

Now we obtain exact closed expressions for the unknown
correlations by using $F1$ and following a method similar to that
provided earlier in the context of kinetic approach for two-phase
turbulent flows\cite{HMR99b}. Considering $\psi[{\bf u}']$ as a
functional of $u_i'$'s having multivariate-normal distribution and
applying $F1$, we obtain
\begin{equation}
\langle \psi({\bf x},t) u_i'({\bf x},t)
\rangle=-\frac{\partial}{\partial x_k}[\lambda_{ki}\langle
\psi\rangle] +\gamma_i \langle \psi \rangle \label{sol1}
\end{equation}
where tensors
\begin{eqnarray}
\lambda_{ki}=\int_{t_0}^t \langle u_i'({\bf
x},t)u_j'(t|t_2)\rangle G_{jk}(t_2|t)dt_2, \label{lambda}\\
\gamma_i=\int_{t_0}^t \Bigl\langle \frac{\partial u_i'({\bf
x},t)}{\partial x_r }u_j'(t|t_2)\Bigr\rangle G_{jr}(t_2|t)dt_2,
\label{gamma}
\end{eqnarray}
and shorthand notation $u_i'(t|t_2)$ is used to represent $u_i'$
at $({\bf X}(t_2),t_2)$ along the fluid particle path which
reaches ${\bf x}$ at $t$. Also $G_{jk}$ is governed by
\begin{equation}
\frac{d}{dt}G_{jk}(t_2|t)-\frac{\partial U_k({\bf x},t)}{\partial
x_a}G_{ja}(t_2|t)=\delta_{jk}\delta({t-t_2})
\end{equation}
and $G_{jk}=\delta_{jk}$ for homogeneous turbulence with uniform
mean velocity $U_k$. The $\lambda_{ki}$ represents eddy
diffusivity and $\gamma_i-\frac{\partial \lambda_{ki}}{\partial
x_k}$ represents drift velocity when Eq. (\ref{sol1}) is
substituted in Eq. (\ref{psi2}). We should mention that for the
homogeneous turbulence with constant $U_k$, the present solution
for $\langle \psi u_i'\rangle$ becomes identical to the solution
derived by Reeks \cite{Reeks01}. Also, the present solution is
valid for both incompressible and compressible velocity fields.
Similarly, considering $\psi_b$ as a functional of $u_i'({\bf
y},s^-)$ and applying $F1$, we obtain
\begin{equation}
\langle\psi_b({\bf y},s^-)u_i'({\bf
y},s^-)\rangle=-\frac{\partial}{\partial
y_k}[\lambda^b_{ki}\langle \psi_b\rangle] +\gamma^b_i \langle
\psi_b \rangle \label{sol1b}
\end{equation}
where the tensors
\begin{eqnarray}
\lambda^b_{ki}=\int_{0}^s \langle u_i'({\bf
y},s^-)u_j'(s^-|s^-_2)\rangle G^b_{jk}(s_2^-|s^-)ds_2,
\label{lambda2} \\
\gamma^b_i=\int_{0}^s \Bigl\langle \frac{\partial u_i'({\bf
y},s^-)}{\partial y_r }u_j'(s^-|s^-_2)\Bigr\rangle
G^b_{jr}(s_2^-|s^-)ds_2,\label{gamma2}
\end{eqnarray}
and $u_j'(s^-|s^-_2)$ represents $u_j'$ at time $s^-_2\equiv
t-s_2$ along the particle trajectory which passes through ${\bf
y}$ at time $s^-\equiv t-s$. Also, $G^b_{jk}(s_2^-|s^-)$ satisfies
\begin{equation}
-\frac{d}{dt}G^b_{jk}(s_2^-|s^-)-\frac{\partial U_k({\bf y
},s^-)}{\partial y_a}G^b_{ja}(s_2^-|s^-)
=\delta_{jk}\delta(s^--s^-_2).
\end{equation}
The right hand side (rhs) of Eqs. (\ref{lambda}), (\ref{gamma}),
(\ref{lambda2}) and (\ref{gamma2}), contain unknown Lagrangian
velocity field correlations $\langle u_i'({\bf
x},t)u_j'(t|t_2)\rangle$, $\langle u_j'(t|t_2){\partial u_i'({\bf
x},t)}/{\partial x_r }\rangle$, $\langle u_i'({\bf
y},s^-)u_j'(s^-|s^-_2)\rangle$ and $\langle
u_j'(s^-|s^-_2){\partial u_i'({\bf y},s^-)}/{\partial y_r
}\rangle$, respectively. The exact equations relating these
correlations to the Eulerian velocity field correlations can be
derived by using the functional formulas. Using $\psi$ and
$\psi_b$, the correlations can be written in the forms
\begin{eqnarray}
\langle a'_{[a']}u_j'(t|t_2)\rangle=\int \langle a'_{[a']}u_j'({\bf
y}_2,s^-_2)\psi_b({\bf y}_2,s^-_2)\rangle d{\bf y}_2, \label{corr1}
\\
\langle c'_{[c']}u_j'(s^-|s^-_2)\rangle=\int \langle
c'_{[c']}u_j'({\bf x}_2,t_2)\psi({\bf x}_2,t_2)\rangle d{\bf x}_2,
\label{l1}
\end{eqnarray}
where $a'_{[a']}\equiv u_i'({\bf x},t)$ or $a'_{[a']}\equiv{\partial
u_i'({\bf x},t)}/{\partial x_r }$ and $c'_{[c']}\equiv u_i'({\bf
y},s^-)$ or $c'_{[c']}\equiv {\partial u_i'({\bf y},s^-)}/{\partial
y_r }$. Here in Eqs. (\ref{corr1}) and (\ref{l1}), we have introduce
another notations $[a']$ and $[c']$ for subscripts and which are
equivalent to the subscripts of selected variables for $a'_{[a']}$
and $c'_{[c']}$. For example, $[a']\equiv i$ and $[a']\equiv ir$
when $a'_{[a']}\equiv u_i$ and $a'_{[a']}\equiv {\partial
u_i}/{\partial x_r}$, respectively. Also, $\psi_b({\bf
y}_2,s^-_2)=\delta({\bf y}_2-{\bf X}(s^-_2))$ and $t_2=t-s_2\equiv
s^-_2$ with initial condition given at time $t$ or $s=0$ and initial
condition for $\psi({\bf x},t_0)=\delta({\bf x}-{\bf y})$ is given
at time $t_0=t-s$. It should be noted that coupling between the
equations for $\langle \psi \rangle$ and $\langle \psi_b \rangle$
arises due to Eqs. (\ref{corr1}) and (\ref{l1}). We now obtain
expressions for third order correlations appearing on the rhs of
Eqs. (\ref{corr1}) and (\ref{l1}).

We introduce a few shorthand notations as $({\bf a })\equiv ({\bf
x},t)$, $({\bf b})\equiv ({\bf y},s^-)$, $({\bf n})\equiv ({\bf
y}_n,s^-_n)$ and $d{\bf n}\equiv d{\bf y}_n\,ds_n$ and
$t_n=t-s_n\equiv s^-_n$ for $n=1,2,\cdots$. Now consider general
form $\langle a'_{[a']}u_j'({\bf y},s^-)\psi_b({\bf y},s^-)\rangle$
with $\psi_b$ as a functional of $u_j'({\bf y},s^-)$ and by applying
$F1, F2$ or $F3, F4$ depending on whether $a'_{[a']}$ is equal to
$u_i'$ or not, we can obtain expression for the third order
correlation as
\begin{equation}
\langle a'_{[a']}({\bf a})u_j'({\bf b})\psi_b({\bf
b})\rangle=\langle a'_{[a']}u_j'({\bf b})\rangle \langle
\psi_b({\bf b})\rangle +{\cal A }_{[a']j}, \label{simp1}
\end{equation}
with
\begin{equation}
{\cal A}_{[a']j}=[\frac{\partial^2}{\partial y_p\partial
y_l}\Lambda_{[a']jpl}-\frac{\partial}{\partial
y_p}\Omega_{[a']jp}-\frac{\partial}{\partial
y_l}\Pi_{[a']jl}+\Gamma_{[a']j}]\langle\psi_b\rangle
\label{sollag1}
\end{equation}
where the tensor
\begin{equation}
\Lambda_{[a']jpl}= \int_0^s ds_1\int_0^s ds_3 \langle
a'_{[a']}u_k'(s^-|s^-_1)\rangle\langle u_j'({\bf
b})u_q'(s^-|s^-_3)\rangle
G_{qp}^b(s_3^-|s^-)G_{kl}^b(s_1^-|s^-)\label{tens1}
\end{equation}
and $(s^-|s^-_n)$ is used to represent value of velocity
fluctuation at $({\bf X}(t-s_n), t-s_n)$ along the fluid particle
trajectory which passes through ${\bf y}$ at time $t-s$. It should
be noted that $0 \le s_n \le s$. The expressions for remaining
tensors $\Omega_{[a']jp}$, $\Pi_{[a']jl}$ and $\Gamma_{[a']j}$ in
Eq. (\ref{sollag1}) are equivalent to the rhs of Eq. (\ref{tens1})
but with changes $u_j'( {\bf b})\rightarrow \frac{\partial u_j'(
{\bf b})}{\partial y_l}$, $u_j'( {\bf b})\rightarrow
\frac{\partial u_j'( {\bf b})}{\partial y_p}$ and $u_j'( {\bf
b})\rightarrow \frac{\partial^2 u_j'( {\bf b})}{\partial
y_p\partial y_l}$, respectively. It should be noted that
substituting only the first term on the rhs of Eq. (\ref{simp1})
into Eq. (\ref{corr1}) would yield results obtained by using
Corrsin's hypothesis \cite{Corrsin59}. The remaining non-zero
terms ${\cal A}_{[a']j}$ and ${\cal B}_{[a']j}$ in Eq.
(\ref{simp1}) account for the correlation between $a'_{[a']},u_j'$
and $\psi_b-\langle \psi_b\rangle$.

Now, the tensors present in Eq. (\ref{sollag1}) contain Lagrangian
correlations of the form $\langle c'_{[c']}u_j'(s^-|s^-_n)\rangle$
and for which expressions can be written similar to that of Eq.
(\ref{l1}). To obtain closed expression for the third order
correlation appearing in Eq. (\ref{l1}), we consider general form
$\langle c'_{[c']}u_j'({\bf a})\psi({\bf a})\rangle$ and by
applying $F1, F2$ or $F3, F4$ depending on whether $c'_{[c']}$ is
equal to $u_i'$ or not, we obtain
\begin{equation}
\langle c'_{[c']}u_j'({\bf a})\psi({\bf a})\rangle=\langle
c'_{[c']}u_j'({\bf a})\rangle\langle\psi ({\bf a})\rangle +{\cal
C}_{[c']j}
\end{equation}
with
\begin{equation}
{\cal C}_{[c']j}=[\frac{\partial^2}{\partial x_p\partial
x_l}\lambda_{[c']jpl}-\frac{\partial}{\partial
x_p}\omega_{[c']jp}-\frac{\partial}{\partial
x_l}\pi_{[c']jl}+\gamma_{[c']j}]\langle\psi\rangle \label{sollag2}
\end{equation}
where the tensor
\begin{equation}
\lambda_{[c']jpl}= \int_{t_0}^t dt_1\int_{t_0}^t dt_3 \langle
c'_{[c']}u_k'(t|t_1)\rangle\langle u_j'({\bf a})u_q'(t|t_3)\rangle
G_{qp}(t_3|t)G_{kl}(t_1|t)\label{tens2}
\end{equation}
 The expressions for remaining
tensors $\omega_{[c']jp}$, $\pi_{[c']jl}$ and $\gamma_{[c']j}$ in
Eq. (\ref{sollag2}) are equivalent to the rhs of Eq. (\ref{tens2})
but with changes $u_j'( {\bf a})\rightarrow \frac{\partial u_j'(
{\bf a})}{\partial x_l}$, $u_j'( {\bf a})\rightarrow
\frac{\partial u_j'( {\bf a})}{\partial x_p}$ and $u_j'( {\bf
a})\rightarrow \frac{\partial^2 u_j'( {\bf a})}{\partial
x_p\partial x_l}$, respectively.  Further, the tensors present in
Eq. (\ref{sollag2}) contain Lagrangian correlations of the form
$\langle a'_{[a']}u_j'(t|t_n)\rangle$, similar to Eq.
(\ref{corr1}) and for which expressions are derived above.

\textit{Concluding Remarks}: In this letter, it has been shown that
a closed set of exact statistical equations describing the turbulent
self-diffusion by multivariate-normal velocity field can be obtained
after solving involved closure problems by using
Furutsu-Donsker-Novikov formula along with other three functional
formulas. The only unknowns in the set are various Eulerian
correlations of the velocity field. We have seen that the
statistical theory of the diffusion requires two scalar fields
$\psi$ and $\psi_b$ describing the forward and backward calculation
of the diffusion with initial conditions given at time $t_0 < t$ and
$t$, respectively, for the evolutions of $\psi$ and $\psi_b$.
Further it has been shown that the Lagrangian correlations appearing
in the theory can be related to Eulerian correlations through the
use of $\psi$ and $\psi_b$ and solving the involved closure problems
in an exact manner by using the functional formulas. The Eulerian
analysis presented in this letter is valid for both homogeneous and
inhomogeneous turbulent velocity field. Lastly, we should mention
that the functional formulas can also be used to derive expressions
for Lagrangian correlations of fluid velocity field along the
inertial particles which occur in the kinetic approach for dispersed
phase of particles/droplets in turbulent flows ( e.g., see Refs.
\cite{PM03} and references cited therein).



\end{document}